\documentclass[11pt]{article}

\usepackage{amsfonts}
\usepackage{amsmath}  
\usepackage{latexsym}
\usepackage{amssymb}
\usepackage{mathrsfs}
\usepackage{tikz}
\usepackage[all]{xy}

\newcommand{\tr}{\text{Tr}}
\newcommand{\ato}[2]{\genfrac{}{}{0pt}{}{#1}{#2}}

\newcommand{\mA}{\mathscr{A}}
\newcommand{\mB}{\mathscr{B}}
\newcommand{\mC}{\mathscr{C}}
\newcommand{\mD}{\mathscr{D}}

\newcommand{\mV}{\mathscr{V}}

\begin{document}

\begin{center}
\textbf{\Large Algebraic Bethe ansatz for the totally asymmetric simple\\ exclusion process with boundaries}\\[2ex]
\large{N. Cramp\'e\footnote{nicolas.crampe@univ-montp2.fr}}\\[2.1ex]
Laboratoire Charles Coulomb (L2C), UMR 5221 CNRS-Univ. Montpellier 2,\\ Montpellier, F-France.
\\[5ex]
 
 \begin{minipage}{13cm}
 \begin{center}
 \textbf{Abstract}
 \end{center}
 \vspace{-3mm}
  \textit{We study the one-dimensional totally asymmetric simple exclusion process in contact with two reservoirs including also a fugacity 
  at one boundary. 
  The eigenvectors and the eigenvalues of the corresponding Markov matrix are computed using the modified algebraic Bethe ansatz, method
  introduced recently to study the spin chain with non-diagonal boundaries. We provide in this case a proof of this method.
  }
 \end{minipage}\vspace{0.5cm}
 \end{center}
MSC numbers (2010): 82B23 ; 81R12

\section*{Introduction}

The one-dimensional totally asymmetric simple exclusion process (TASEP),
describing the diffusion of particles with hard-core interactions, is one
of the most studied models of non-equilibrium statistical mechanics (see \cite{Der,Sch} for reviews).
The stationary state of this model has been computed in
\cite{DEHP} using the method so-called matrix ansatz (see \cite{Eva} for a review). 
Then, a link with the integrable spin chains has been found in \cite{sandow} allowing one to use the results
obtained in the context of the integrable systems to study the TASEP.
For example, the algebraic Bethe ansatz has been used in 
\cite{DE} to compute the spectral gap of diffusion models using the previous works on the integrable quantum spin chain \cite{CLSW,nepo}.

For the computation of the current fluctuations, the situation is more complicated. 
The matrix ansatz has only been developed recently in \cite{GLMV1} for the TASEP then generalized 
in \cite{GLMV2} for the ASEP. The use of the algebraic Bethe ansatz was only possible for a discrete set of the parameters of the ASEP \cite{GE}. 
The main result of this paper is to provide the algebraic Bethe ansatz for any parameters of the TASEP using the recent results 
introduced in the context of the spin chains \cite{BC,Bel}. Let us mention that the results of the papers \cite{BC,Bel} have been conjectured by investigating
models of small size. Then, the result of \cite{BC} has been proven in \cite{ZLCYSW}. The results of this paper are proven and therefore
support the conjectures of \cite{Bel}. The proofs proposed here could be also useful to study other models. 

The plan of this paper is as follows. In section \ref{sec:tasep}, we present the model and give the eigenvalues of the generalized Markov 
matrix as well as the associated Bethe equations. Then, we introduce the transfer matrix associated to the TASEP in section \ref{tra}
and we give the outline of the modified algebraic Bethe ansatz in section \ref{sec:maba}. More technical proofs are given in section \ref{sec:pr}.

\section{Totally asymmetric simple exclusion process and its eigenvalues\label{sec:tasep}}

\begin{figure}[htb]
\begin{center}
 \begin{tikzpicture}[scale=0.7]
\draw (-2,0) -- (9,0) ;
\foreach \i in {-2,-1,...,9}
{\draw (\i,0) -- (\i,0.4) ;}
\draw[->,thick] (-2.4,0.9) arc (180:0:0.4) ; \node at (-2.,1.8) [] {$\alpha$};
\draw  (1.5,0.5) circle (0.3) [fill,circle] {};
\draw  (4.5,0.5) circle (0.3) [fill,circle] {};
\draw  (5.5,0.5) circle (0.3) [fill,circle] {};
\draw  (8.5,0.5) circle (0.3) [fill,circle] {};
\draw[->,thick] (1.6,1) arc (180:0:0.4); \node at (2.,1.8) [] {$1$};
\draw[->,thick] (4.6,1) arc (180:0:0.4);
\draw[thick] (4.8,1.7) -- (5.2,1.1);\draw[thick] (5.2,1.7) -- (4.8,1.1);
\draw[->,thick] (8.6,1) arc (180:0:0.4) ; \node at (9.,1.8) [] {$\beta$};
 \end{tikzpicture}
 \end{center}
 \caption{TASEP model.}
 \label{fig:tasep}
\end{figure}
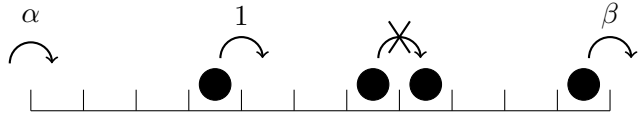

We consider the TASEP on a finite segment of size $L$ in contact with two reservoirs.
The dynamics of the model is defined by the following rules: each site can be occupied by at most one particle, a particle attempts to hop 
on its right neighboring site with rate $1$ unless this site is occupied, a particle may appear at the site $1$ with rate $\alpha$ if this site is empty 
and a particle may disappear at the site $L$ with rate $\beta$ (see figure \ref{fig:tasep}). 

A configuration of the system is given by a $L$-tuple $(\tau_1,\tau_2,\dots,\tau_L)$
where $\tau_i=1$ if a particle is present at the site $i$ and $\tau_i=0$ otherwise. The probabilities of each configuration 
at time $t$, $P_t(\tau_1,\tau_2,\dots,\tau_L)$,
can be encompassed in the following vector 
\begin{equation}
 P_t=\left(
 \begin{array}{c}
  P_t(0,\dots,0,0,0 )\\
   P_t(0,\dots,0,0,1 )\\
   P_t(0,\dots,0,1,0 )\\
   \vdots\\
   P_t(1,\dots,1,1,1 )
 \end{array}
\right)=
\sum_{\tau_1,\tau_2,\dots,\tau_L=0,1}P_t(\tau_1,\tau_2,\dots,\tau_L)\ e^{\tau_1}\otimes e^{\tau_2}\otimes\dots\otimes e^{\tau_L}\;,
\end{equation}
where $e^0=\left(\begin{array}{c}
                    1\\0
                   \end{array}\right)$
                   and
$e^1=\left(\begin{array}{c}
                    0\\1
                   \end{array}\right)
$. Then, the time evolution of the probability is given by the following master equation
\begin{equation}
 \frac{d P_t}{dt}=MP_t\;.
\end{equation}
The Markov matrix, $M$, for the TASEP is given by
\begin{equation}
 M=B_1 +\sum_{k=1}^{L-1} w_{k,k+1}  +\widetilde B_L\;,
\end{equation}
where the subscripts indicate on which sites the matrices $w$, $B$ and $\widetilde B$ act on non trivially and
\begin{equation}
 B=\left(\begin{array}{c c}
    -\alpha & 0\\
    \alpha & 0
   \end{array}\right)
\quad,\quad
w=\left(\begin{array}{c c c c}
    0 & 0 & 0 & 0\\
    0 & 0 & 1 & 0\\
    0 & 0 & -1 & 0\\
     0 & 0 & 0 & 0
    \end{array}\right)
\quad,\quad
 \widetilde B=\left(\begin{array}{c c}
    0 & \beta\\
    0 & -\beta
   \end{array}\right)\;.
\end{equation}
For the boundaries rates, we will use also the convenient notations $a=\frac{1}{\alpha}-1$ and $b=\frac{1}{\beta}-1$.

To compute the fluctuations of the current, an off-diagonal element of the Markov matrix is multiplied by a fugacity $e^\mu$ 
to keep track of the number of particles jumping through a bond \cite{LS}. 
We choose here to count the number of particles entering in the system and the corresponding generalized Markov matrix is 
\begin{equation}\label{eq:gm}
 M(\mu)=B_1(\mu) +\sum_{k=1}^{L-1} w_{k,k+1}  +\widetilde B_L\qquad\text{where} \quad B(\mu)=\left(\begin{array}{c c}
    -\alpha & 0\\
    \alpha e^\mu & 0
   \end{array}\right)\;.
\end{equation}
Its largest eigenvalue is the generating function of the cumulants of the current in the long time limit.

The main result of this paper is to compute the eigenvalues and the eigenvectors of $M(\mu)$ with the algebraic Bethe ansatz.
Before to give details concerning the computations of the eigenvectors, we summarize the results about the eigenvalues.
The eigenvalues of the generalized Markov matrix are
\begin{equation}\label{eq:l}
 \lambda=-\beta-\sum_{p=1}^L\frac{u_p}{u_p-1}\;,
\end{equation}
where $u_1,\dots,u_L$ are solutions of the Bethe equations
\begin{equation}\label{eq:be}
(au_j+e^\mu)\ \left(\frac{(u_j-1)^2}{u_j}\right)^L=(au_j+1)\ (u_j+b)\ \prod_{\ato{k=1}{k\neq j}}^{L}
 \left(u_j-\frac{1}{u_k}\right)\ ,\qquad\text{for $j=1,2,\dots,L$}\;.
\end{equation}
These results are a direct consequence of the general results of section \ref{sec:maba}.
The eigenvalues \eqref{eq:l} are given by
\begin{equation}\label{eq:lL}
\lambda=-\frac{1}{2}
\left.\frac{d \Lambda(x)}{dx}\right|_{x=1,z_i=1}
\end{equation}
where $\Lambda(x)$ is the eigenvalues of the transfer
matrix \eqref{eq:L}. The Bethe equations \eqref{eq:be} are obtained by
setting $z_i=1$ in \eqref{eq:beinh}. 

As usual, we consider only the solutions of the Bethe equations such that the Bethe roots are two by two different. 
We solved numerically Bethe equations \eqref{eq:be} for systems of small size ($L=1,2,3$) and we compared with a direct 
diagonalisation of the generalized Markov matrix. We showed that, in these cases, the spectrum obtained by the Bethe 
equations is complete.

The comparison of our result with previous results may be fruitful: Bethe equations \eqref{eq:be} must be 
a limit of the Bethe equations obtained for the XXZ spin chain \cite{CYSW2}, Bethe vectors \eqref{eq:vec} are conjectured in \cite{Bel}
for the XXZ spin chain and the eigenvalue \eqref{eq:l} with the largest real part must be compared to the one obtained with matrix ansatz \cite{GLMV1,GLMV2}.

\paragraph{Markovian model.}
The Markovian model is recovered for $e^\mu=1$.
In this case, Bethe equations \eqref{eq:be} split into two cases:
\begin{itemize}
\item For $u_j\neq -1/a$ ($j=1,\dots,L$),  
the factors $(au_j+1)$ can be simplified on both sides of the Bethe equations to transform them into
\begin{equation}\label{eq:bel}
 \left(\frac{(u_j-1)^2}{u_j}\right)^L=(u_j+b)\  \prod_{\ato{k=1}{k\neq j}}^{L}
 \left(u_j-\frac{1}{u_k}\right)\qquad\text{for $j=1,2,\dots,L$}\;.
\end{equation}
By solving these Bethe equations \eqref{eq:bel} for small size systems, we show that they seem to have only one solution 
corresponding to the stationary state of the TASEP (\textit{i.e.} with vanishing eigenvalue $\lambda=0$). 
Although the result for the eigenvalue is very simple, it seems that there are no simple expressions for the Bethe roots (see Fig. \ref{fig:bethe} 
for an example).
\begin{figure}[htb]
\begin{center}
\includegraphics[trim=0 15cm 7cm 0,scale=0.6]{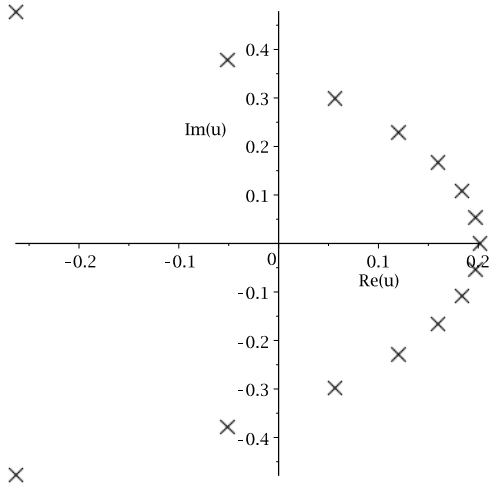} 
\caption{Bethe roots in the complex plane, solutions of \eqref{eq:bel} for $L=15$ and $b=0.8$. \label{fig:bethe}}
\end{center}
\end{figure}
It would be very interesting to compare 
the results obtained here and the matrix ansatz \cite{DEHP}.
\item Since all the Bethe roots must be distinct, we may choose without loss of generality $u_L=-1/a$  which is a solution of the 
$L^\text{th}$ Bethe equation\footnote{We may choose any other Bethe root equals to $-1/a$ but, by invariance of the Bethe equations by permutations, 
we recover the same solutions.}. The $L-1$ remaining Bethe equations become
\begin{equation}\label{eq:bel2}
 \left(\frac{(u_j-1)^2}{u_j}\right)^L=  (u_j+b)\ \left(u_j+a\right) \prod_{\ato{k=1}{k\neq j}}^{L-1}
 \left(u_j-\frac{1}{u_k}\right)\qquad\text{for $j=1,2,\dots,L-1$}\;.
\end{equation}
The associated eigenvalues can be written as $\displaystyle \lambda=-\alpha-\beta-\sum_{p=1}^{L-1}\frac{u_p}{u_p-1}$.
The Bethe equations \eqref{eq:bel2} have been used previously in \cite{DE} to compute the spectral gap. They show that
all the eigenvalues except the stationary state are obtained. 
\end{itemize}
In conclusion, for $e^\mu=1$, the complete spectrum is obtained by solving Bethe equations \eqref{eq:bel} and \eqref{eq:bel2}.

\section{Transfer matrix and algebraic Bethe ansatz \label{sec:tr_maba}}

\subsection{Transfer matrix \label{tra}}

As usual in the context of the algebraic Bethe ansatz, one diagonalizes, instead of
the Markov matrix, the transfer matrix.
The central objects to construct the transfer matrix are the R-matrix, solution of the Yang-Baxter equation, 
and the K-matrix, solution of the reflection equation (see \cite{CRV} for a review about the transfer matrix for the exclusion processes). 
For the TASEP, they are given explicitly by
\begin{equation}\label{eq:R}
R(x)=\left( \begin {array}{cccc} 
1&0&0&0\\ 
0&0&x&0\\
0&1&1-x&0\\
0&0&0&1
\end {array} \right)\ ,\
K(x)=\left( \begin {array}{cc} 
\displaystyle{\frac{(a+x)x}{x a+1}}&0\\[2ex]
\displaystyle{e^\mu\ \frac{1-x^2}{x a+1}}&1
\end {array} \right)
\ \text{and}\ \
\widetilde K(x)= \frac{1}{xb+1}\left( \begin {array}{cc} 
1&1\\ 
0&xb
\end {array} \right)\;.
\end{equation}
Then, one defines the monodromy matrix by
\begin{equation}\label{eq:B}
 B_{0}(x)= R_{0L}(x/z_L) \cdots R_{01}(x/z_1)\ K_{0}(x)\  R_{10}(xz_1) \cdots R_{L0}(xz_L)
 =:\left(\begin{array}{c c}
       \mA(x) & \mB(x)\\
       \mC(x) & \mD(x)
      \end{array}
\right)\;
\end{equation}
where $z_1,\dots,z_L$ are called inhomogeneity parameters and
the transfer matrix by \cite{sklyanin}
\begin{equation} 
t(x) = \tr_{0}\ \big(\widetilde K_{0}(x)\ {B}_{0}(x)\big) =\frac{1}{xb+1}\left(\mA(x)+xb\mD(x)+\mC(x)\right)\,.
\label{transfer}
\end{equation}
The important features of the transfer matrix are that they commute for different spectral parameters\footnote{In the case 
of the TASEP model, the proof given in \cite{sklyanin} cannot be used directly 
because the crossing symmetry is not satisfied by the R-matrix \eqref{eq:R}. However, it is valid for
the partially asymmetric simple exclusion process and then we can take the limit to get the result for the TASEP \cite{CRV}. }
(\textit{i.e.} $[t(x),t(y)]=0$)
and that the generalized Markov matrix is obtained by 
\begin{equation}\label{eq:dtM}
 -\frac{1}{2}
\left.\frac{d t(x)}{dx}\right|_{x=1,z_i=1}=M(\mu)\;.
\end{equation}
Then, the eigenvectors of the generalized Markov matrix $M(\mu)$ can be computed by putting $z_i=1$
in the eigenvectors of the transfer matrix. 

The monodromy matrix satisfies also the reflection equation and one deduces that 
\begin{eqnarray}
&&[\mC(x),\mC(y)]=0\ \text{,}\qquad\mD(x)\mC(y)=\frac{x(xy-1)}{y-x}\mC(y)\mD(x)-\frac{y(xy-1)}{y-x}\mC(x)\mD(y)-\mC(x)\mA(y)\;.\label{eq:DC}
\end{eqnarray}
Unfortunately, there exists no relation allowing us to move $\mA$ from the left to the right of $\mC$ which complicates our tasks: 
we will come back to this point in section \ref{sec:prAD}. This feature is particular to the TASEP and is due to the $0$ on the 
diagonal of the R-matrix \eqref{eq:R}. We can also show that $\mB(x)=0$.\\

To conclude this section, we would like to mention that the problem to find exact methods to solve problem 
with non-diagonal boundaries (\textit{i.e.} $K$ and $\widetilde K$ are not diagonal) 
has attracted a lot of attention. The problem lies in the fact that the usual methods are based on the 
existence of one simple particular eigenvector which does not exist in this case.
Therefore, numerous approaches have been modified and generalized to deal with this problem:
the algebraic Bethe ansatz \cite{CLSW,MMR,YanZ07,BCR13,pimenta}, the functional Bethe ansatz \cite{nepo,MurN05,Gal08,FraGSW11}, 
the coordinate Bethe ansatz \cite{CRS1}, the separation of variables \cite{FSW,niccoli2}, 
the q-Onsager approach \cite{BK} and the matrix ansatz \cite{GLMV1,GLMV2,CRS2}.
Recently, inhomogeneous T-Q relations have been studied in \cite{CYSW2,CYSW1,Nep} where they obtained eigenvalues and Bethe equations for generic boundaries.
These results have permitted to conjecture a modified algebraic Bethe ansatz to get the eigenvectors \cite{BC,Bel} (proved in the case of the open XXX chain 
in \cite{ZLCYSW}). It is this last method we used in this paper to find the eigenvalues and the eigenvectors of the generalized Markov matrix.

\subsection{Modified algebraic Bethe ansatz \label{sec:maba}}

The modified algebraic Bethe ansatz states that
eigenvectors of the transfer matrix are given by a product of $L$ matrices $\mC$ where $L$ is the number of sites of the model.
Therefore, eigenvectors of $t(u_0)$ are given by the Bethe vector
\begin{equation}\label{eq:vec}
 \Phi(u_1,u_2,\dots,u_L)=\mC(u_1)\mC(u_2)\dots \mC(u_L)|\Omega\rangle\,,
\end{equation}
where $|\Omega\rangle=e^1\otimes e^1\otimes \dots \otimes e^1$ and $\{u_1,u_2,\dots,u_L\}$ are solutions of Bethe equations:
\begin{equation}\label{eq:beinh}
(au_j+e^\mu)\ \prod_{\ell=1}^L\frac{(u_j-z_\ell)(u_jz_\ell-1)}{u_jz_\ell}=  (u_j+b)\ (au_j+1)\  \prod_{\ato{k=1}{k\neq j}}^{L}
 \left(u_j-\frac{1}{u_k}\right)\ ,\qquad\text{for $j=1,2,\dots,L$}\;.
\end{equation}
The associated eigenvalues are
\begin{equation}\label{eq:L}
 \Lambda(u_0)=u_0^{L+1}\frac{b+u_0}{bu_0+1} \prod_{k=1}^L \frac{u_0u_k-1}{u_k-u_0}
 -\frac{(au_0+e^\mu)(u_0^2-1)}{(au_0+1)(bu_0+1)}\prod_{j=1}^L\left[(u_0-z_j)(u_0-\frac{1}{z_j})\right] \prod_{k=1}^L \frac{u_k}{u_k-u_0}\;.
\end{equation}
Let us emphasize that the main difference with the usual algebraic Bethe ansatz is that the number of matrices $\mC$
must be equal to the number of sites of the model. A vector constructed with less than $L$ matrices $\mC$ (in particular $|\Omega\rangle$) 
is not an eigenvector of the transfer matrix. Another difference is the presence of the matrix $\mC$ in the transfer matrix and the necessity to compute
its action on $\Phi$: it is this computation which forces us to take $L$ matrices $\mC$ in $\Phi$ (see section \ref{sec:prC}).
The rest of the paper is devoted to prove these results. As usual in the context of the algebraic Bethe ansatz, the completeness 
of such solution is not proven but it is conjectured supported by numerical evidences (as explained previously in section \ref{sec:tasep}).

As for the usual algebraic Bethe ansatz, we need the actions of the matrices $\mA$ and $\mD$ 
on a product of $\mC$. In the section \ref{sec:prAD}, we show the following relations
\begin{eqnarray}
&&\hspace{-2cm} \mA(u_0) \Phi(u_1,\dots,u_L)=
\sum_{p=0}^L\frac{u_0u_p(u_p^2-1)}{u_pu_0-1}\prod_{\ato{k=0}{k\neq p}}^L\left( \frac{u_p(u_pu_k-1)}{u_k-u_p} \mC(u_k)\right)|\Omega\rangle\nonumber\\
&& \hspace{1.4cm}-
 \sum_{p=0}^L\frac{au_p(u_p^2-1)}{au_p+1}\prod_{j=1}^L\left[(u_p-z_j)(u_p-\frac{1}{z_j})\right]\prod_{\ato{k=0}{k\neq p}}^L\left( \frac{u_k}{u_k-u_p} \mC({u}_k)\right)|\Omega\rangle\label{eq:A}\\
 &&\hspace{-2cm}\mD(u_0) \Phi(u_1,\dots,u_L)=
 \sum_{p=0}^L\frac{u_p^2-1}{u_pu_0-1}\prod_{\ato{k=0}{k\neq p}}^L\left( \frac{u_p(u_pu_k-1)}{u_k-u_p} \mC({u}_k)\right)|\Omega\rangle\;.\label{eq:D}
\end{eqnarray}
Due to the presence of the operator $\mC$ in the transfer matrix, we need also the following relation 
proven in section \ref{sec:prC}
\begin{equation}
 \mC(u_0) \Phi(u_1,\dots,u_L)=
 e^\mu\sum_{p=0}^L\frac{1-u_p^2}{au_p+1} \prod_{j=1}^L\left[(u_p-z_j)(u_p-\frac{1}{z_j})\right] \prod_{\ato{k=0}{k\neq p}}^L
 \left( \frac{u_k}{u_k-u_p} \mC({u}_k)\right)|\Omega\rangle\;.\label{eq:C}
\end{equation}

Now, we are in position to compute the action of the transfer matrix on $\Phi(u_1,u_2,\dots,u_L)$ :
\begin{eqnarray}
 t(u_0)\Phi(u_1,\dots,u_L)&=&\frac{1}{bu_0+1}\big(\mA(u_0)+bu_0\mD(u_0)+\mC(u_0)\big)\ \Phi(u_1,\dots,u_L)\\
 &=&\Lambda(u_0)\Phi(u_1,u_2,\dots,u_L) + \sum_{p=1}^{L} F(u_0,u_p) U_p \prod_{\ato{k=0}{k\neq p}}^L \mC(u_k)|\Omega\rangle\label{eq:O}
\end{eqnarray}
where $F(u,x)=\frac{u(x^2-1)}{(u-x)(bu+1)}$
and
\begin{equation}
U_p=(b+u_p)u_p^L\prod_{\ato{k=1}{k\neq p}}^L\frac{u_pu_k-1}{u_k-u_p}
-\frac{au_p+e^\mu}{au_p+1} \prod_{j=1}^L\left[(u_p-z_j)(u_p-\frac{1}{z_j})\right]  \prod_{\ato{k=1}{k\neq p}}^L\frac{u_k}{u_k-u_p}\;.
\end{equation}
Relation \eqref{eq:O}, called off-shell equation, is obtained using relations \eqref{eq:A}, \eqref{eq:D} and \eqref{eq:C} and particularizing the elements 
$p=0$ in the sum. Bethe equations \eqref{eq:beinh} imply the vanishing of $U_p$ (for $p=1,2,\dots,L$) and we obtain 
that $\Phi(u_1,u_2,\dots,u_L)$ is an eigenvector of $t(u_0)$
with the eigenvalue $\Lambda(u_0)$.

\section{Actions of $\mA$, $\mC$ and $\mD$ on the Bethe vector $\Phi$ \label{sec:pr}}

In the previous section \ref{sec:maba}, we gave the outline of the modified algebraic Bethe ansatz but the 
central relations \eqref{eq:A}, \eqref{eq:D} and \eqref{eq:C} are only proven in this section. These proofs are more 
technical and we prefer, for clarity, to write them separately.

\subsection{Proof of relation \eqref{eq:C} \label{sec:prC}}

Relation \eqref{eq:C} is a new type of relation to prove in comparison to the usual algebraic Bethe ansatz.
To demonstrate it, let us introduce the following vector
\begin{equation}\label{V}
 \mV(x)=e^\mu \frac{1-x^2}{x(ax+1)}\prod_{j=1}^L\left[(x-z_j)(x-\frac{1}{z_j})\right] 
 \prod_{\ell=0}^L\left( \frac{u_\ell}{u_\ell-x}\mC(u_\ell)\right)  \mC(x)^{-1} |\Omega\rangle\;.
\end{equation}
We are going to show that the entries of this vector have only poles at $x=0$ and $x=u_p$ ($p=0,1,\dots,L$).

Firstly, we perform a change of basis using the factorizing twist introduced in \cite{MSS} to obtain a simple explicit formula for $\mC(x)$. 
The factorizing twist is
\begin{equation}
 F_{12\dots L}=F_{L-1,L}F_{L-2,L-1L}\dots F_{1,23\dots L}
\end{equation}
where
\begin{equation}
 F_{j,j+1\dots L}=F_{j,j+1\dots L}(z_j,\dots,z_L)=1-\hat n_j +\hat n_j R_{jL}(z_j/z_L)\dots R_{jj+1}(z_{j}/z_{j+1})\;.
\end{equation}
We have introduced the matrix $\hat n=\left(\begin{array}{c c}
                                       0 &0\\
                                       0&1
                                      \end{array}\right)
$. Using the results of \cite{MSS}, one gets
\begin{eqnarray}
 &&\mC^F(x)=F_{12\dots L}\mC(x)F^{-1}_{12\dots L}=\frac{1-x^2}{ax+1}
 \Bigg[\sum_{i=1}^L x(z_i+a)\sigma^+_i\prod_{\ato{j=1}{j\neq i}}^L\left( (1-xz_j)\hat n_j+\frac{x-z_j}{z_i-z_j}(1-\hat n_j)\right) \nonumber\\
 &&\hspace{6.5cm}+e^\mu\prod_{j=1}^L\left( (1-xz_j)\hat n_j+\frac{z_j-x}{z_j}(1-\hat n_j)\right)
 \Bigg]\;.\label{c}
\end{eqnarray}

Secondly, by noting that $\mC^F(x)$ is an upper triangular matrix, we can compute the determinant of $\mC(x)$:
\begin{equation}
 \det(\mC(x))=\det(\mC^F(x))=\left(e^\mu \frac{1-x^2}{ax+1}\right)^{2^L}\prod_{j=1}^L\left( (1-xz_j)\frac{z_j-x}{z_j}\right)^{2^{L-1}}\;.
\end{equation}
For a generic value of $x$, the determinant does not vanish which allows us to take the inverse of $\mC^F(x)$ (and also of $\mC(x)$ which 
justifies the definition of $\mV(x)$). 

Thirdly, we can determine the entries of $\mC^F(x)^{-1} |\Omega\rangle$ 
\begin{equation}\label{cinv}
\mC^F(x)^{-1} |\Omega\rangle =\frac{(ax+1)e^{-\mu}}{(1-x^2)\prod_{j=1}^L\left[(x-z_j)(x-\frac{1}{z_j})\right] }\ \sum_{\epsilon_1,\dots,\epsilon_L=0,1}\ 
 \prod_{j=1}^L f_{\epsilon_j}(x,z_j) \ \  e^{\epsilon_1}\otimes\dots\otimes e^{\epsilon_L}\;,
\end{equation}
where $f_\epsilon(x,z)=(\epsilon-1)e^{-\mu}(a+z)x +\epsilon (z-x)/z$. We demonstrate relation \eqref{cinv} by showing that, with this expression and 
expression \eqref{c} of $\mC^F$, we get $\mC^F(x)\mC^F(x)^{-1} |\Omega\rangle=|\Omega\rangle$.

Finally, we remark that $F_{12\dots L}\mV(x)$ is equal to the R.H.S. of \eqref{V} replacing all the $\mC$ by $\mC^F$ 
(since $F_{12\dots L} |\Omega\rangle=|\Omega\rangle$ \cite{MSS}) and
we deduce that $F_{12\dots L}\mV(x)$ has only poles
at $x=0$ and $x=u_p$ ($p=0,1,\dots,L$) since from \eqref{V} and \eqref{cinv}, we show that there 
are no poles at $x=1,-1,-1/a,z_j,1/z_j,\infty$. Let us remark that, if we take less than $L+1$ matrices $\mC$ in the definition of $\mV$, there 
is a pole at infinity and the corresponding residue does not take a nice form.

Therefore, the only non trivial residues of $\mV(x)$ are
\begin{eqnarray}
 &&\left. \text{Res}\ \mV(x)\right|_{x=0}=e^\mu \prod_{l=0}^L \mC(u_l)  \mC(0)^{-1} |\Omega\rangle\\
 &&\left. \text{Res}\ \mV(x)\right|_{x=u_p}=-e^\mu\frac{1-u_p^2}{au_p+1} \prod_{j=1}^L\left[(u_p-z_j)(u_p-\frac{1}{z_j})\right] \prod_{\ato{k=0}{k\neq p}}^L
 \left( \frac{u_k}{u_k-u_p} \mC({u}_k)\right)|\Omega\rangle\;.
\end{eqnarray}
By using that $\mC(0)^{-1} |\Omega\rangle=F_{12\dots L}^{-1}\mC^F(0)^{-1} |\Omega\rangle=e^{-\mu} |\Omega\rangle$ and that the sum over all the residues of a 
rational function vanishes, we prove relation \eqref{eq:C}.

\subsection{Proof of relations \eqref{eq:A} and \eqref{eq:D} \label{sec:prAD}}

As mentioned in section \ref{tra}, the relations of type \eqref{eq:A} and \eqref{eq:D} are usually proven using the commutation 
relation between $\mA$, $\mD$ and $\mC$. Unfortunately in the case of the TASEP model, no relation permuting $\mA$ and $\mC$ exists.
To overcome this problem, we use the transfer matrix associated to the Partially Asymmetric Simple Exclusion Process (PASEP)
depending on the parameter $q$ such that we recover the TASEP in the limit $q\rightarrow 0$.

The R-matrix associated to the PASEP is given by
\begin{equation}\label{eq:RASEP}
R^{(q)}(x)=\left( \begin {array}{cccc} 
1&0&0&0\\ 
0&\frac{(x-1)q}{qx-1}&\frac{(q-1)x}{qx-1}&0\\
0&\frac{q-1}{qx-1}&\frac{x-1}{qx-1}&0\\
0&0&0&1
\end {array} \right) .
\end{equation}
One gets $R^{(0)}(x)=R(x)$ where $R(x)$ is the R-matrix of the TASEP \eqref{eq:R}. We indicate by the superscript $(q)$ the objects defined in section \ref{tra}
but for the R-matrix $R^{(q)}$. From now, the computations are similar to the ones one performs usually in the context of the algebraic Bethe ansatz.

The reflection equation satisfied by the monodromy matrices $B^{(q)}(x)$ allows us to get the following commutation relations
\begin{eqnarray}
 \overline\mA^{(q)}(x)\mC^{(q)}(y)&=&\frac{(q^2xy-1)(qx-y)}{q(x-y)(qxy-1)}\ \mC^{(q)}(y)\overline\mA^{(q)}(x)
 -\ \frac{(q-1)(q^2xy-1)x}{q(x-y)(qx^2-1)}\ \mC^{(q)}(x)\overline\mA^{(q)}(y)\nonumber\\
 &&+\ \frac{xy(q-1)(y^2-1)(q^2x^2-1)}{(qxy-1)(qx^2-1)(qy^2-1)}\ \mC^{(q)}(x)\mD^{(q)}(y)\label{qAC}\\
 \mD^{(q)}(x)\mC^{(q)}(y)&=&\frac{(x-qy)(xy-1)}{(qxy-1)(x-y)}\mC^{(q)}(y)\ \mD^{(q)}(x)+\ \frac{(q-1)(y^2-1)y}{(x-y)(qy^2-1)}\ \mC^{(q)}(x)\mD^{(q)}(y)\nonumber\\
 &&-\ \frac{q-1}{qxy-1}\ \mC^{(q)}(x)\overline\mA^{(q)}(y)\label{qDC}
\end{eqnarray}
where $\overline\mA^{(q)}(x)=\mA^{(q)}(x)+\frac{(1-q)x^2}{qx^2-1}\mD(x)$. We see that relation \eqref{qDC} gives back relation \eqref{eq:DC} in the limit 
$q\rightarrow 0$ whereas relation \eqref{qAC} is not defined in this limit. 

We can also determine the values of 
$\overline\mA^{(q)}(x)$ and $\mD^{(q)}(x)$ on the vector $|\Omega\rangle$ and we get
\begin{eqnarray}
 &&\overline\mA^{(q)}(x)\ |\Omega\rangle=q^L\ \frac{x(x^2-1)(qx+a)}{(xa+1)(qx^2-1)}\ \prod_{j=1}^{L}\frac{(z_j-x)(1-xz_j)}{(z_j-qx)(1-qxz_j)}\ |\Omega\rangle\\
 &&\mD^{(q)}(x)\ |\Omega\rangle=|\Omega\rangle
\end{eqnarray}
By using these previous relations, we are able to compute $\mA^{(q)}(u_0) \Phi^{(q)}(u_1,\dots,u_L)$ and $\mD^{(q)}(u_0) \Phi^{(q)}(u_1,\dots,u_L)$. 
The results are not singular in the limit $q\rightarrow 0$ and we get relations \eqref{eq:A} and \eqref{eq:D}.

\paragraph{Acknowledgement:} I thank warmly S.Belliard, V.Caudrelier, E.Ragoucy and M.Vanicat for their interests and their suggestions.

\end{document}